\documentclass{webofc}
\usepackage[varg]{txfonts}   
\usepackage{listings}
\usepackage{hyperref}
\begin{document}
\title{San Pedro Meeting on Wide Field Variability Surveys:\\ Some Concluding Comments.} 
%
%

\author{\firstname{Michael W.} \lastname{Feast}\inst{1,2}\fnsep\thanks{\href{mailto:mwf@ast.uct.ac.za}{\tt mwf@ast.uct.ac.za}} 
}

\institute{Astronomy Dept.,University of Cape Town 
\and
South African Astronomical Observatory}

\abstract{
  This is a written version of the closing talk at the 22nd Los Alamos Stellar pulsation conference on wide field variability surveys. It comments on some of the issues which arise from 
 the meeting. These include the need for attention to photometric standardization (especially in the infrared) and the somewhat
controversial problem of statistical bias in the use of parallaxes (and other methods of distance determination). Some major
advances in the use of pulsating variables to study Galactic structure are mentioned. The paper includes a clarification of
apparently conflicting results from
classical Cepheids and RR Lyrae stars in the inner Galaxy and bulge. The importance of understanding non-periodic phenomena in variable stars,
particularly AGB variables and RCB stars is stressed, especially for its relevance to mass-loss, in which pulsation may only play
a minor role.
}
\maketitle

\section{Introduction}\label{intro}
This has been a meeting full of interest. In part, this is due to the range of topics discussed which reflect the overlapping areas in which large scale variability surveys are having a major
impact. The traditional pulsators (RR Lyraes, Cepheids, Miras etc) remain of major importance for our understanding of stellar structure and evolution. But they are also increasingly important elements in our understanding of the structure and evolution of our own and nearby galaxies as well as being a key to the cosmic distance scale. 

The main interest of the founders of these meetings can be
correctly described as asteroseismology. The field to which this term now applies has made remarkable advances for our understanding of stars. Currently this new asteroseismology is expanding
into the area of Galactic structure by giving us ages, masses and luminosities of large numbers of stars. One sees a bright future for both the traditional pulsators and the new asteroseismology
in Galactic studies. But we need to remember that in such work, we are not alone. Major surveys such as Apogee are revealing much about the composition and structure of our Galaxy without any
reference to variability and these will need to be considered together with the wide field variability results.

Rather than trying to summarize all the topics discussed at this meeting, I would like simply to mention a few issues which arise, as well
as a relevant area which has only been mentioned briefly.
\section{Some Technical Issues}\label{tech}
\subsection{Automated Classification}\label{auto}
Automated classification and machine learning were discussed in an invited review. Obviously we want to divide variable stars into physically
homogeneous groups. But how this is related to observed features such as light curve shapes, remains a learning process and we need to be
prepared for surprises. For instance Pietrzy{\'n}ski et al.\cite{pietrzynski} find that an OGLE star classed as an RR Lyrae variable is in fact an object
in a binary with a mass of only 0.26 solar masses. Again, are we correct in placing the metal rich and metal poor RR Lyraes together; for instance,
in a single (visual) luminosity sequence (a point raised by Chadid et al.\cite{chadid})? 
Published classifications from light curves have often been based on relatively limited material
and some types of variable are then rather difficult to distinguish from one another. For instance only the accurate photometry and densely sampled
light curves of the OGLE survey allowed a few classical Cepheids in the extreme outer flared disc of our Galaxy to be correctly identified and studied \cite{feast1}.
\subsection{Photometric Calibration}\label{photometry}
 It is sometimes said that large scale photometric surveys are self-calibrating and do
not need to tie themselves to other photometric systems. There is some truth in this. However, 
in the case of variable star surveys, it is to be hoped that efforts will be made to link the results
with other commonly used systems. This is for at least two reasons. First, for at least the brighter variables,
other observations are likely to exist. Secondly, in deep surveys the brightest variables will
probably be saturated. These stars will need to be included in a variety of analyses. One such, key,
place is luminosity calibration. We can expect to have absolute magnitudes in various common systems for various variability
classes, e.g. from parallaxes of bright variables, and these will have to be converted into the adopted survey system.
The alternative is to make special observations of bright variables in the same system as the deep survey. 
Linking space-base photometry with ground-based work is a particular problem and is one that is crucial to
cosmic distance scale work. The project to observe Milky Way classical Cepheids with good parallaxes directly with
the HST for a distance ladder calibration is therefore of considerable importance
(\cite{reiss} and the talk by Lucas Macri).

A related problem is the level of of consistency between different observers within one photometric system. It is perhaps not
always realized how much effort went into the standardization of the $UBVRI$ system. Despite work by various groups there has been,
so far as I know, no such massive effort for infrared systems. For instance in $JHK$ (or $K_{s}$) different groups use slightly different filters
and different standard stars. In this situation, the transformation between systems (including transformation of reddening laws)  is often not well
known. Such matters are particularly important in distance scale applications.  A warning example is the local group dwarf galaxy NGC 6822. Two groups
obtained infrared photometry of classical Cepheids in this galaxy. For 11 Cepheids in common the differences are: $\Delta J = 0.126\pm 0.022$
and $\Delta K_s = 0.061 \pm 0.014$\cite{feast2}. The reason for these differences is not known, but they may be at least
partly due to transformation problems. Differences of this size in a nearby galaxy obviously raise warning flags if we
are attempting to obtain $H_0$ to 2.4 percent  ($\pm 0.05$mag ) e.g. Lucas Macri's talk.   
\subsection{Parallaxes}\label{parallaxes} 
\subsubsection{GAIA}\label{gaia}
Gisella Clementini gave us a very clear talk on the work she and her colleagues have been doing on the Gaia parallaxes
of classical Cepheids and RR Lyrae variables. The Gaia results are very spectacular and will become more so with later
data releases. Gisella made some comparison with the pulsation parallaxes (Baade-Wesselink method) of RR Lyrae variables 
from Muraveva et al.\cite{muraveva} (the derived parallaxes can be obtained from the data they publish). I also made this 
comparison. The results are really quite
remarkable. For 19 RR Lyraes in common between Gaia and Muraveva, the unweighted mean difference is only
$0.02 \pm 0.03$mas. The scatter about the mean is only $0.14$mas, whilst the unweighted mean Gaia standard error (s.e.) is
twice as much, $0.28$mas. This suggests that the Gaia workers are perhaps being overcautious in their assignments of 
standard errors. However, when applying the Gaia results to Galactic structure and other
problems, well determined standard errors will often be require, if only for assessing any statistical bias. 
Indeed, Gisella mentioned that different methods of estimating bias gave different results. It will obviously be useful
to consider these in  detail when the full results are published, together with considerations such as give in 2.3.2 below.

\subsubsection{Statistical Bias}\label{bias}
Statistical bias needs to be considered in many problems involving distance estimates whether using parallaxes or other
methods. Methods of dealing with distance, or distance related, bias have a long history going back to Eddington \cite{eddington1} \cite{eddington2}.
However, the literature shows that there remain misunderstandings and uncertainties in this area. 

 An interesting and
relevant example is in the papers by Benedict et al.\cite{benedict1}\cite{benedict2} which deal with the calibration of the absolute magnitudes
of classical Cepheids and RR Lyrae variables using HST parallaxes. The authors of these papers (of which I was one)
could not agree on the question of bias corrections and the method to be used in the analyses. The results for two methods
were therefore given in the papers. One of these (denoted as A below) involves a Lutz-Kelker (L-K) type correction as recently restated by Benedict et al. \cite{benedict3} and 
this solution was used in a recent $H_0$ calibration 
 by Reiss et al.\cite{reiss}. The other (denoted as B below) takes into account only a (very small) Malmquist-type bias correction.

The problem rests with the manner in which the stars were chosen for observation by the HST and also how the resulting parallaxes are used.
In the present case the RR Lyraes and Cepheids were chosen, as the nearest suitable examples of these classes, on the basis of their
photometric distance moduli (or relative moduli). In the case of the Cepheids it so happened that a selection according to {\it Hipparcos} parallaxes
gave the same sample. However, these {\it Hipparcos} parallaxes have significant uncertainties while those of the relative photometric parallaxes are small. Thus
it is best to consider the samples as photometrically selected\footnote {It should be noted that even if we did consider a {\it Hipparcos} selected sample
it would be the {\it Hipparcos} uncertainties that would be relevant in any correction for selection bias, not the HST ones.
Such corrections would in the present case be large and very uncertain.}. It is clear that if the
photometric parallaxes had no uncertainties (i.e. were the true parallaxes) 
the HST parallaxes would be distributed about them in an unbiased manner
(first equation in the Lutz-Kelker paper \cite{lutz} (see e.g. \cite{feast6}) and this argument can be generalized.
The application of a  L-K type bias correction to the HST data is therefore not appropriate. However samples chosen by photometric parallax will be biased with
respect to a volume limited sample and thus, in principle, require a Malmquist-type correction, though in the present cases this is small.

Method A also combines the data in magnitude space which gives rise to biased weighting since this involve $\pi/\sigma_\pi$
and overweights/ underweights accidentally too large/ too small parallax measures. 
Method B avoids this by working in  parallax space \cite{feast3} \cite{feast4}.
Fortunately in the cases being discussed,
the resulting difference between the two analyses was small; 
the different relative weightings offsetting
the difference due to a correction or otherwise for supposed L-K effect.

As a general comment it is perhaps worth remarking that, though L-K bias and other statistical corrections may sometimes be appropriate,
it is not always recognized that, if these corrections are substantial, their uncertainties can be large and highly asymmetric. Estimates
of these uncertainties, in the appropriate situations, can be taken from Koen \cite{koen} if one takes his {\it N} to be the number of stars
in the sample, each with a parallax standard error of {\it s}.

\section{Galactic Structure}\label{Gal}
This meeting has shown how large photometric surveys, extending from the optical to the infrared, are changing our views
of our own Galaxy. This is especially true if optical, infrared, spectroscopic and radial velocity data can be combined.
New surveys and Gaia data will lead, one can be sure, to further major advances.

 Recent variability work has
extended from the central nuclear disc to the outer fringes of the halo. The following are  brief comments on
some of these results, some of which require further extensive follow up work.
\subsection{Nuclear Stellar Disc}\label{Nuclear}
In the centre of the galactic bulge, i.e. in the nuclear stellar disc, Matsunaga et al.(\cite{matsunaga1}\cite{matsunaga2} and this meeting) have discovered a few
classical Cepheids. The lack of short period Cepheids, indicates that star formation rapidly increased there about
25Myr ago and this activity continued until quite recently. These Cepheids were found using the 1.4m Japanese-South African Infrared Survey Facility (IRSF).
We can expect the powerful VISTA telescope to tell us much more about this important region of the Galaxy which is hidden
by dust at optical wavelengths.
\subsection{The Galactic bulge}\label{bulge}
Recent years have seen a vast amount of work on the Galactic bulge. This has shown the bulge together with its bar (or bars) to
be much more complex than was at one time thought. Some of the variable star work in this area has been covered in talks at this meeting
(by, e.g., Soszy{\'n}ski, Minniti and Matsunaga). One importance of variable stars for work of this kind is the possibility of separating
populations according to age. An example of this is the recent work of Catchpole et al. \cite{catchpole} which examined the distribution
of bulge Miras as a function of period. This is an indication of the their distribution according to age and shows that  the longer period, intermediate age
objects populate a bar.   

A recent work of particular interest to this meeting is
the VISTA survey for RR Lyrae variables towards the Galactic centre (\cite{minniti} and Dante Minniti this meeting). The VISTA field
covers about 1.5 square degrees though much of this is very heavily obscured. It is suggested that the 14 RR Lyrae stars found are distinct
from the general RR Lyrae bulge population and come from globular clusters destroyed in the very high density regions near the centre.
This is an interesting possibility. However, even if these stars are at the distance of the centre, distinguishing clearly
between the central condensation of the general RR Lyrae population and a separate ``globular-derived" group will be difficult. The general
bulge RR Lyrae population is expected to be high near the centre since even well out in the bulge the optical OGLE survey finds an average of
171 RR Lyraes per sq.deg. \cite{soszynski}. The question of whether these RR Lyraes are close to the Centre or possibly somewhat nearer
is discussed briefly at the end of subsection 3.3 below.
\subsection{Cepheids and the Extreme Inner Disc}\label{Cepheids}
A search for classical Cepheids in the general direction of the Galactic centre has been made by two groups,Matsunaga et al.(\cite{matsunaga3} and this meeting)
using the IRSF,
and D{\'e}k{\'a}ney et al. (\cite{dekany1}\cite{dekany2} and this meeting) using VISTA. There is reasonable agreement with the Cepheids discovered and,
within the limits set by transformations of photometric systems, their photometry.
However, the interpretation the two groups give to the data differs, and this difference is of considerable importance for our understanding of
the structure and evolution of the inner Galactic disc. 

These Cepheids are all highly reddened and much of the difference between the two groups
depends on the adopted reddening law (\cite{nishiyama1} (=N6), in the case of Matsunaga et al.{\cite{matsunaga3}, and \cite{nishiyama2} (=N9) in the case
of D{\'e}k{\'a}ny et al.\cite{dekany2}.  
I can only briefly sketch out my current understanding of this matter which 
is from the point of view of a member of the Matsunaga group (which includes Nishiyama).

N6 is based on red clump stars, observed with the IRSF,
close to the Galactic centre. N9 is based on red clump stars together with red giants from 2MASS transformed by D{\'e}k{\'a}ny et al. into the VISTA system.
Although the difference  between the two reddening laws is not highly significant, it is important in these highly reddened regions. N6 was preferred
by Matsunaga et al. partly on the general grounds that the law was derived entirely from IRSF observations as were the Cepheid observations. 
There are also empirical reasons for adopting N6.
Matsunaga et al.\cite{matsunaga1}\cite{matsunaga2} give good reasons for placing the "nuclear disc Cepheids" (see section 3.1 above) close to the centre which adopting the N6 reddening law does. 
N9 would place these stars nearer to us. Moreover, there is a 3kpc region
essential devoid of Cepheids around these central stars. This is present independent of the adopted reddening law. It would be off-centred with N9.
Such a distribution would be very hard to understand. This "Cepheid-free" region could be due to the sweeping effect of the Galactic bar or perhaps
the 3kpc expanding arm. 

In a paper at this meeting, present on his behalf by G. Hajdu, I. D{\'e}k{\'a}ny argued that a {$J-H/H-K$ plot favoured N9. This matter
will be considered in detail elsewhere. However, it should be noted that neither set of workers use the J band data for distances. Also, it is not clear
whether, for instance, the colour-colour relation derived is consistent with Cepheid intrinsic colours or whether the result is statistically significant. It was also argued that the RR Lyraes in the 
direction of the centre\cite{minniti} (section 3.2 above) would be closer than the Centre if N6 is used. However, these stars would
be even nearer to us if N9 had been used. So the argument is not relevant in distinguishing between N6 and N9. 
Minitti et al.\cite{minniti} place their RR Lyraes near the centre using an estimate of the reddening law by
Alonso-Garc{\'i}a et al.\cite{alonso}. The Alonso-Garc{\'i}a result is an order of magnitude less certain than N6 and the difference from N6 is not highly significant.
In fact the large number of RR Lyraes that are expected in the direction of the centre together with the fact that the VISTA survey in this direction is J magnitude limited,
indicates that  the Vista RR Lyrae stars \cite{minniti} could well be in the foreground. More work on this point would clearly be useful.       
\subsection{Halo and Outer Disc}\label{halo}
Large scale surveys have played a major role in revealing the structure of the outer regions of our Galaxy and its environment. This work continues,
for instance at this meeting Judith Cohen presented velocity data on Halo RR Lyraes beyond 50kpc and the extensive work on RR Lyraes
in stellar streams in and beyond the Halo is well known. Classical Cepheids in the flared outer disc were mentioned in section 2.1.
\section{Variable stars and mass loss}\label{massloss}
\subsection{AGB variables}\label{AGB}
Mass loss from stars is important for stellar evolution, the contents of the interstellar medium, Galactic chemical evolution
and dust in the Universe. Much of this mass loss come from variable stars near the top of the AGB and this has generally
been attributed to the initial ejection of matter by stellar pulsations, followed by the formation of grains and  their further
ejection by radiation pressure. Yet, perhaps rather
strangely we have heard rather little about mass loss at this meeting. Possibly this is partly due to the embarrassment that adopting 
this hypothesis has led to little or no success in obtaining a quantitative explanation of
the phenomenon. It has been clear for many years (e.g. Whitelock et al.\cite{whitelock1}) that for large amplitude AGB variables (Miras)
there is a good correlation between period, pulsational amplitude and dust shell mass. More recent studies (e.g. Goldman et al.\cite{goldman}) get similar results.
But it is perhaps time to ask whether these empirical relations are telling us any more than that AGB pulsators can get very large
and very cool thus facilitating mass loss and dust formation. Do the correlations tell us anything about the physical causes of mass loss
from large, cool, stars? 
Evidence has been around for some while that though outward
velocities associated with pulsations may aid mass loss, other major factors are at work. For instance C-rich Mira variables in our own
and nearby galaxies (e.g. Whitelock et al.\cite{whitelock2} \cite{whitelock3}) are found to go through protracted "obscuration events"  due to dust (soot) formation. 
In cases where suitable studies have been made (e.g. \cite{feast5}), the events are found to be due to dust clouds ejected in the line
of sight and are not spherically symmetrical ejections.
 
These ejected clouds are similar to the dust puff events in RCB stars and like them
have been attributed to ejection from large turbulent convection cells. Support for this conclusion has come,
in more recent times,  from the  very beautiful work by a number of groups who have carried out high resolution work or interferometry on AGB variables.
For instance the work of Ohnaka et al.\cite{ohnaka} shows the complex structures of gas and dust round the oxygen-rich Mira, W Hya. Similar results have been obtained
for other mass-losing AGB variables (e.g. \cite{haubois}, \cite{wittkowski}).
3-D radiation-hydrodynamic techniques \cite{hofner1} \cite{hofner2} are beginning to model these outer structures
and asymmetries. Work along these lines may eventually enable AGB mass loss to be understood quantitatively. Nevertheless there 
are formidable problems to solve. For instance, what is presumably require is not just some mean understanding of turbulent convection
but extreme values. This is analogous to the problem of convective overshooting in stellar interiors. 
The frequency and size of ejected dust clouds is not understood. Declines can last for many years
e.g. in the case of a recent decline of R For, 18 years \cite{feast5}. More observations would be helpful to get proper statistics and this seems a programme 
for small automated survey telescopes covering large areas to only moderate depth. These lower cost projects have perhaps the best
chance of lasting for the very many years required. 
\subsection{R Coronae Borealis type variables}\label{RCB}
One mass loss topic that was discussed at this meeting was that of the RCB variables. A good survey  of the RCB problem in general
was give by Geoff Clayton. This included a summary of the evidence that pulsation is the cause of mass ejection which is known to be
in puffs or clouds. Considerations such as those just mentioned in the case of AGB stars can be raised regarding pulsation as a physical
mechanism for RCB mass loss. Indeed the evidence of any link of pulsation to mass loss is not, in my view strong. This is certainly the
case for the type star R CrB. Roger Griffin has been monitoring this star for many years with a {\it Coravel} type radial velocity
machine and other instruments. This work together with early work by Herbig and covering more than 50 years (with gaps) is being prepared
for publication. One point, relevant to the cause of mass loss, comes from Griffin's observations at the time of the great decline which began in 2007
and still continues. During a period of about 100 days prior to the decline, (part of) the stellar atmosphere was highly disturbed with high velocity
outflows ($\rm \sim 40\, km.s^{-1}$). It is believed that these outflows eject mass to sufficient distance for dust to form. Dust formation and declines, though
spectacular, are simply secondary phenomena of physical processes (probably connected to turbulent convection) on the stellar surface. 
As with the AGB stars the frequency of occurrence and lengths of RCB declines is not understood.
\section{Summary}\label{summ} 
 Current surveys of pulsating variables are already revealing much new about our own and nearby galaxies
and much more can be expected from forthcoming surveys and their follow-up. Asteroseismology will soon be doing
the same, 
as well as telling us about stellar structure. Surveys of classical pulsators will lead to a fuller
physical understanding of them and provide an increasingly firmer base for their use in cosmological and other
studies. The reasons for adopting a Galactic disc model with a deficit of classical Cepheids in the inner regions are set out.
To make full use of present and future large scale variability surveys attention needs to be given to the problems
of photometric standardization. The paper attempts to clarify and settle a problem related to the analysis of HST
parallaxes of classical Cepheids and RR Lyrae stars.

Large scale surveys will lead to a better
understanding of non -periodic phenomena in, for instance, AGB variables. Understanding these phenomena and their relation
to mass loss is of importance both for stellar and galactic evolution.


\begin{acknowledgement} 
\noindent\vskip 0.2cm
\noindent {\em Acknowledgments}: I am very grateful to my colleagues, Noriyuki Matsunaga, Patricia Whitelock and John Menzies 
for many helpful discussions and comments. The work was supported by a research grant to MWF from the South African National Research Council (NRF).
\end{acknowledgement}


\begin{thebibliography}{}
\bibitem{alonso}
Alonso-Garc{\'i}a, J., D{\'e}k{\'e}ny, I., Catelan, M., Ramos, R.C., Gran, F., Amigo, P., Leyton, P., Minniti, D., AJ, {\bf 149}, 99 (2015)
\bibitem{benedict1}
Benedict, G.F., MacArthur, B.E., Feast, M.W., Barnes, T.G., Harrison, T.E., Paterson, R.J., Menzies, J.W., Bean, J.L., et al.
AJ, {\bf 133}, 1810 (2007)
\bibitem{benedict2}
Benedict, G.F., MacArthur, B.E., Feast, M.W., Barnes, T.G., Harrison, T.E., Bean, J.L., Menzies, J.W., Chaboyer, B., et al., AJ, {\bf 142}, 187 (2011)                                                                                                                                                   
\bibitem{benedict3}
Benedict, G.F., McArthur, B.E., Nelan, E.P., Harrison, T.E., PASP, {\bf 129}, 012001 (2017)
\bibitem{catchpole}
Catchpole, R.M., Whitelock, P.A., Feast, M.W., Hughes, S.M.G., Irwin, I., Alard, C., MNRAS, {\bf 455}, 2216 (2016) 
\bibitem{chadid}
Chadid, M., Sneden, C., Preston, G.W., arXiv:1611.02368
\bibitem{dekany1}
D{\'e}k{\'e}ny, I, Minniti, D., Hajdu, G., Alonso-Garc{\'i}a, J., Hempel, M., Palma, T., Catelan, M., Gieren, W., Majaess, D.,
ApJL, {\bf 799}, L1 (2015)
\bibitem{dekany2}
D{\'e}k{\'e}ny, I, Minniti, D., Majaess, D., Zoccali. M., Hajdu, G., Alonso-Garc{\'i}a, J., Catelan, M., Gieren, W., Borissova, J.,
ApJL, {\bf 812}, L29 (2015)
\bibitem{eddington1} 
Eddington, A.S. MNRAS, {\bf 73}, 359 (1913)
\bibitem{eddington2}
Eddington, A.S. \textit{Stellar Motions and the Structure of the Universe}, p.~172 (MacMillan, London, 2014)
\bibitem{feast6}
Feast, M.W., in, \textit{Globular Clusters} , p.~251 (Cambridge University Press), ed. Mart\'{i}nez Roger, Fourn\'{o}n, S\'{a}nchez (1999)
\bibitem{feast3}
Feast, M.W., MNRAS, {\bf 337}, 1035 (2002)
\bibitem{feast5}
Feast, M.W., Whitelock, P.A. \& Marang, F., MNRAS, {\bf 346}, 878 (2003)
\bibitem{feast2}
Feast, M.W.,  Whitelock, P.A., Menzies, J.W. \& Matsunaga, N., MNRAS, {\bf 421}, 2998 (2012)
\bibitem{feast4}
Feast, M.W., in \textit{Planets, Stars and Stellar Systems}, vol.5, (Springer, Dordrecht), ed. T.D. Oswalt \& G. Gilmore p.~829 (2013)
\bibitem{feast1}
Feast, M.W., Menzies, J.W., Matsunaga, N. \& Whitelock, P.A., Nature, {\bf 509}, 342 (2014)
\bibitem{hofner1}
Freytag, B., \&  H{\"o}fer, S., A\&A, {483}, 571 (2008)
\bibitem{goldman}
Goldman, S.R., van Loon, J.Th., Zijlstra, A.A., Green,J.A., Wood, P.R., Nanni, A., Matsuura, M., Groenewegen, M.A.T., et al. MNRAS, {\bf 465}, 403 (2017) 
\bibitem{hanson}
Hanson, R.B., MNRAS, {\bf 186},875 (1979)
\bibitem{haubois}
Haubois, X., Wittkowski, M., Perrin, G., Kervella, P., M{\'e}rand, A., Th{\'e}baut, E., Ridgway, S.T., Ireland, M. et al. A\&A, {\bf 582}, 71 (2015)
\bibitem{hofner2}
 H{\"o}fer, S., arXiv:1610.08937
\bibitem{koen}
Koen, C., MNRAS, {\bf 256},65 (1992)
\bibitem{lutz}
Lutz, T.E., \& Kelker, D.H., PASP, {\bf 85}, 573 (1973) 
\bibitem{matsunaga1}
Matsunaga, N., Kawadu, T., Nishiyama, S., Nagayama, T., Kobayashi, N., Tamura, M., Bono, G., Feast, M.W., et al. Nature, {\bf 477}, 188 (2011)
\bibitem{matsunaga2}
Matsunaga, N., Fukue, K., Yamamoto R., Kobayashi, N., Inno, L., Genovali, K., Bono, G., Baba.J., et al. ApJ, {\bf 799}, 46 (2015)
\bibitem{matsunaga3}
Matsunaga, N., Feast, M.W., Bono, G., Kobayashi, N., Inno, L., Nagayama, T., Nishoyama, S., Matsuoka, Y., et al., MNRAS, {\bf 462}, 414, (2016)
\bibitem{minniti}
Minniti, D., Ramos, R.C., Zoccali, M., Rejkuba, M., Gonzalez, O.A., Valenti, E., \& Gran, F., ApJl, {\bf 830}, L14 (2016)
\bibitem{muraveva}
Muraveva, T., Palmer, M., Clementini, G., Luri, X., Cioni, M-R.L., Moretti, M.I., Marconi, M., Riperi, V. et al. ApJ, {\bf 807}, 127 (2015)
\bibitem{nishiyama1}
Nishiyama, S., Nagata, T., Kusakabe, N., Matsunaga, N., Naoi, T., Kato, D., Nagashima, C., Sugitani, K., et al., ApJ, {\bf 638} 839 (2006)( =N6)
\bibitem{nishiyama2}
Nishiyama, S., Tamura, M., Hatano, H., Kato, D., Tanab{\'e}, T., Sugitani, K., Nagata, T., ApJ, {\bf 696}, 1407 (2009)(=N9)
\bibitem{ohnaka}
Ohnaka, K., Wiegert., G \& Hofmann, K-H., A\&A {\bf 589}, 91 (2016)
\bibitem{pietrzynski}
Pietrzy{\'n}ski, G., Thompson, I.B., Gieren, W., Graczyk, D., Stepie{\'n}, K., Bono, G., Moroni, P.G.P., Pilecki, B., et al. Nature, {\bf 484}, 76 (2012)
\bibitem{reiss}
Riess, A.G., Macri, L.M., Hoffmann, S.L., Scolnic, D., Casertano, S., Filippenko, A.V., Tucker, B.E., Reid, M.J. et al.,  ApJ, {\bf 826}, 56 (2016)
\bibitem{soszynski}
Soszy{\'n}ski, I., Dziembowski, W.A., Udalski, A., Poleski, R., Szyma{\'n}ski, M.K., Kubiak, M., Pietrzy{\'n}ski, G., Wyrzykowski, \L., et al. Act. Ast. {\bf 61}, 1 (2011)
\bibitem{whitelock1}
Whitelock, P.A., Pottasch, S.R., \& Feast, M.W., in \textit{Late Stages of Stellar Evolution} (Reidel, Dordrecht) ed. S Kwok \& S.R. Pottasch, p.~269 (1987) 
\bibitem{whitelock2}
Whitelock, P.A., Feast, M.W., Marang, F., \& Groenewegen, M.A.T., MNRAS, {\bf 369}, 751 (2006) 
\bibitem{whitelock3}
Whitelock, P.A., Menzies, J.W., Feast, M.W., Matsunaga, N., Tanab{\'e}, T., Ita, Y., MNRAS, {\bf 394}, 795 (2009) 
\bibitem{wittkowski}
Wittkowski, M., Hofmann, K.-H., H{\"o}fner, S., Le Bouquin, J.B., Nowotny, W., Paladini, C., Young, J., Berger, J.-K. et al., arXiv:1702.02574
\end{thebibliography}
%
%

\end{document}